\documentclass[prd,aps,eqsecnum,showpacs]{revtex4}

\usepackage{amsmath,amssymb,amsthm,graphicx,latexsym}

\usepackage{pstricks}
\usepackage{color}

\usepackage{mathrsfs}
\usepackage{hyperref}

\newcommand{\be}{\begin{equation}}
\newcommand{\ee}{\end{equation}}
\newcommand{\bea}{\begin{eqnarray}}
\newcommand{\eea}{\end{eqnarray}}
\newcommand{\bml}{\begin{subequations}}
\newcommand{\eml}{\end{subequations}}
\newcommand{\bfig}{\begin{figure}}
\newcommand{\efig}{\end{figure}}

\begin{document}

\title{Brane inflation in background supergravity}

\author{Sayantan Choudhury$^{1}$\footnote{Electronic address: {sayanphysicsisi@gmail.com}} ${}^{}$
and Supratik Pal$^{1, 2}$\footnote{Electronic address:
{pal@th.physik.uni-bonn.de}} ${}^{}$}
\affiliation{$^1$Physics and Applied Mathematics Unit, Indian Statistical Institute, 203 B.T. Road, Kolkata 700 108, India \\
$^2$Bethe Center for Theoretical Physics and Physikalisches
Institut der Universit\"{a}t Bonn, Nussallee 12, 53115 Bonn,
Germany}


\begin{abstract}

We propose a model of  inflation in the framework of brane
cosmology driven by background supergravity. Starting from bulk
supergravity we construct the inflaton potential on the brane and
employ it to investigate for the consequences to inflationary
paradigm. To this end, we derive the expressions for the important
parameters in brane inflation, which are somewhat different from
their counterparts in standard cosmology, using the one loop radiative corrected 
potential. We further  estimate the observable parameters
and find them to fit well with  recent observational data by confronting with WMAP7 using CAMB. We
also analyze the typical energy scale of brane inflation with our
model, which resonates well with present estimates from cosmology
and standard model of particle physics.

\end{abstract}

\pacs{98.80.-k ; 98.80.Cq ; 04.50.-h}

\maketitle


\section{\bf INTRODUCTION}

Investigations for the crucial role of Supergravity  in explaining
cosmological inflation date back to early eighties of the last
century (for two exhaustive reviews see \cite{nills} and
 \cite{Antonio} and references therein). One of the generic features of the inflationary paradigm
based on SUGRA is the well-known $\eta$-problem, which appears in
the F-term inflation due to the fact that the energy scale of
F-term inflation is induced by all the couplings via vacuum energy
density.  Precisely, in the expression of F-term inflationary
potential a factor $\exp\left(K/M_{PL}\right)$ appears, leading to
the second slow roll parameter $\eta \gg 1$, thereby violating an
essential condition for slow roll inflation. The usual wayout is
to impose additional symmetry to the framework. One such symmetry
is Nambu-Goldstone shift symmetry \cite{Yanagida} under which
K$\ddot{a}$hler metric becomes diagonal which serves the purpose
of canonical normalization and stabilization of the volume of the
compactified space. Consequently, the imaginary part of the scalar
field gives a flat direction leading to a successful model of
inflation. An alternative approach is to apply noncompact
Heisenberg group transformations of two or more complex scalar
fields where one can exploit Heisenberg symmetry \cite{Gaillard}
to solve $\eta$-problem.  The role of  K\"{a}hler geometry to solve $\eta$-problem in the context of
N=1 SUGRA under certain constraints can be found in \cite{Laura}.

Of late the idea of braneworlds came forward \cite{rs}. From
cosmological point of view the most appealing feature of brane
cosmology is that the 4 dimensional Friedmann equations are to
some extent different from the standard ones due to the
non-trivial embedding in the $S^{1}/Z_{2}$ orbifold
\cite{Maartens}. This opens up new perspectives to look at the
nature in general and cosmology in specific. To mention a few, the
role of the projected bulk Weyl tensor appearing in the modified
Friedmann equations has been studied extensively for metric-based
perturbations \cite{pertmet}, density perturbations on large
scales  \cite{pertden},
  curvature perturbations  \cite{pertcur} and
Sachs-Wolfe effect \cite{brsachs}, vector perturbations
\cite{pertvec}, tensor perturbations \cite{pertten} and CMB
anisotropies \cite{pertcmb}. Brane inflation in the above
framework has also been studied to some extent \cite{hime,
Bassett,  A.A.Sen}. Apart from these phenomenological approaches,
some other approaches which are more appealing in dealing with
fundamental aspects such as possible realization in string theory
can be found in \cite{henrytye, selftune, lambda, axion}. For
example, an apparent conflict between self-tuning mechanism and 
volume stabilization has been shown in \cite{selftune}, subsequently, this problem
has been resolved in \cite{lambda} where
the credentials of the dilatonic field in providing a natural explanation for dark energy by an effective scalar field
on the brane has been demonstrated using self-tuning mechanism in (4+2) dimensional bulk space time.
 The role of the axions as quintessential
candidates has been revealed in \cite{axion}.

In the Randall-Sundrum two-brane scenario \cite{rs} where
the bulk is five dimensional with the fifth dimension
compactified on the orbifold $S^{1}/Z_{2}$ of comoving radius R,
the separation between the two branes give rise to a field -- the
so-called {\it radion} -- which plays a crucial role in governing
dynamics on the brane. The well-known Goldberger-Wise mechanism
\cite{goldberger} leading to several interesting ideas deal with
different issues related to radion. Subsequently, in order to incorporate
observationally constraint cosmology of the brane, a fine tuning between the brane tension of the visible and invisible brane 
has been proposed \cite{extra}. It has been pointed out in
\cite{Georg, Lahanas} how the radion coupled with bulk fields may
give rise to an effective inflaton field on the brane. In the same
vein, we construct the  brane inflaton potential of our
consideration starting from  5D SUGRA. 
In brane inflation   the modified Friedman equations lead to a
modified version of the slow roll parameters \cite{Maartens}.
So, by construction, $\eta$-problem is smoothened to some extent 
by modification of Friedmann equations on the brane \cite{dvali, A.A.Sen}. In a sense, this is a parallel
approach to the usual string inflationary framework where $\eta$-problem is resolved by
fine-tuning  \cite{shamit}. As it will appear, there is still some fine-tuning required in brane inflation, which
arises via a new avatar of five-dimensional Planck mass  
but it is softened to some extent due to the modified Friedman equations.

 As we will find in the present article the proposed model
of brane inflation  matches quite well with latest observational
data from WMAP \cite{wmap07} and is expected to fit well with
upcoming data from Planck \cite{planck}. To this end, we
explicitly derive the expressions for different observable
parameters from our model and further estimate their numerical
values finally leading to confrontation with observation using
the publicly available code CAMB \cite{camb}. We have also analyzed
the typical energy scale of brane inflation and found it to be in
good agreement  with present estimates of cosmological frameworks
as well as standard model of particle physics.


\section{\bf Modeling brane inflation}

Let us consider an effective $N=1, D=4$ SUGRA inflationary potential in the brane derived from $N=2, D=5$ SUGRA in
the bulk. How we have arrived at an effective $N=1, D=4$ SUGRA in the brane starting from $N=2, D=5$ SUGRA in
the bulk and the subsequent form of the loop corrected potential stated in eqn(\ref{opdsa}) has been discussed in details in the Appendix.
For convenience, let us express the one loop corrected renormalizable
potential in terms of inflationary parameters  as
\be\label{opdsa}
V(\phi)=\Delta^{4}\left[1+\left(D_{4}+K_{4}\ln\left(\frac{\phi}{M}\right)\right)
\left(\frac{\phi}{M}\right)^{4}\right],\ee where we introduce new
constants defined by ( $C_4$ is negative in tree level) 
$K_{4}=\frac{9\Delta^{4}C^{2}_{4}}{2\pi^{2}M^{4}} , ~~
D_{4}=C_{4}-\frac{25K_{4}}{12}.$
It is the Coleman Weinberg potential \cite{Coleman}
,\cite{Landau} , provided the coupling constant satisfies the
Gellmann-Low equation in the context of Renormalization group
\cite{das},\cite{ryder}. Here the first term  is constant and
 physically represents the energy scale of inflation ($\Delta$).

\begin{figure}[htb]
{\centerline{\includegraphics[width=7cm, height=5cm] {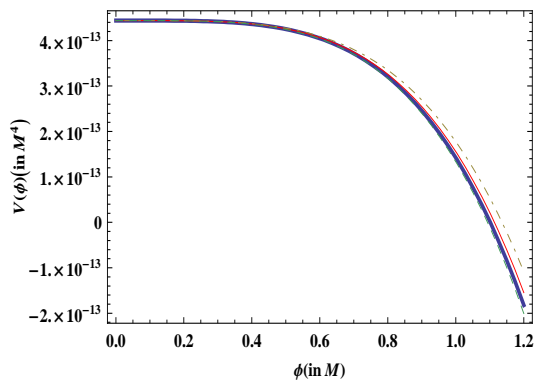}}}
\caption{Variation of one loop corrected potential($V(\phi)$) versus inflaton field ($\phi$)} \label{figVr845}
\end{figure}

Figure (\ref{figVr845}) represents the inflaton potential for
different values of $C_{4}$, $D_{4}$ and $K_{4}$. From the observational constraints
the best fit model is given by the range $-0.70<D_{4}<-0.60$  so
that while doing numericals we shall restrict ourselves to this
range of $D_{4}$. In what follows our primary intention will be to
engage ourselves in modeling brane inflation and to search for its
pros and cons with the above potential (\ref{opdsa}). We shall
indeed find that brane inflation with such a potential
successfully explains the CMB observations
and thus leads to a promising model of inflation.

As already mentioned, the most appealing feature of brane
cosmology is that the 4 dimensional Friedmann equations are to
some extent different from the standard ones due to the
non-trivial embedding in the $S^{1}/Z_{2}$ manifold
\cite{Maartens}. At high energy
regime one can neglect the contribution from Weyl term and consequently, the brane Friedmann
equations are given by \cite{Maartens, Ellwanger} 
$H^{2}=\frac{8\pi
V}{3M^{2}_{PL}}\left[1+\frac{V}{2\lambda}\right]$. The modified
Freidmann equations, along with the Klein Gordon equation, lead to
new slow roll conditions  and new
expressions for observable parameters as well \cite{Maartens,Ellwanger}.
 For convenience throughout the analysis we define the following global functions of the inflaton field
\be\begin{array}{lllll}\label{h341} L(\phi)=\left[1+\frac{\alpha}{2}S(\phi)\right],~~~~
 T(\phi)=\left[1+\alpha S(\phi)\right],~~
 S(\phi)=\left[1+\{D_{4}+K_{4}\ln\left(\frac{\phi}{M}\right)\}
\left(\frac{\phi}{M}\right)^{4}\right],\\
U(\phi)=\left[(K_{4}+4D_{4})+4K_{4}\ln\left(\frac{\phi}{M}\right)\right],~~ 
E(\phi)=\left[(7K_{4}+12D_{4})+12K_{4}\ln\left(\frac{\phi}{M}\right)\right],\\
F(\phi)=\left[(26K_{4}+24D_{4})+24K_{4}\ln\left(\frac{\phi}{M}\right)\right],~~
J(\phi)=\left[(50K_{4}+24D_{4})+24K_{4}\ln\left(\frac{\phi}{M}\right)\right],\\
\tilde{\bar{P}}(\phi)=\sqrt{\left[1+2\alpha S(\phi)L(\phi)\right]}
-2\alpha S(\phi)L(\phi)\sinh^{-1}\left(\left[2\alpha S(\phi)L(\phi)\right]\right)^{-1/2}
\end{array}\ee with $\alpha=\Delta^{4}/\lambda$. 
Incorporating the potential of our consideration from
Eq (\ref{opdsa}) the
 slow roll parameters turn out to be
\bea \label{first} \epsilon_{V}
&=&\frac{M^{2}_{PL}}{16\pi}\left(\frac{V^{'}}{V}\right)^{2}\frac{1+\frac{V}{\lambda}}{(1+\frac{V}{2\lambda})^{2}}
=
\frac{U^{2}(\phi)T(\phi)}{2S^{2}(\phi)L^{2}(\phi)}\left(\frac{\phi}{M}\right)^{6},
\\
\label{second}\eta_{V} &=&
\frac{M^{2}_{PL}}{8\pi}\left(\frac{V^{''}}{V}\right)\frac{1}{(1+\frac{V}{2\lambda})}
=\frac{E(\phi)}
{S(\phi)L(\phi)}\left(\frac{\phi}{M}\right)^{2},
\\
\label{third} \xi_{V} &=&
\frac{M^{4}_{PL}}{(8\pi)^{2}}\left(\frac{V^{'}V^{'''}}{V^{2}}\right)\frac{1}{(1+\frac{V}{2\lambda})^{2}}
=\frac{U(\phi)F(\phi)}
{S^{2}(\phi)
L^{2}(\phi)}\left(\frac{\phi}{M}\right)^{4},
\\
\label{fourth} \sigma_{V} &=&
\frac{M^{6}_{PL}}{(8\pi)^{3}}\frac{(V^{'})^{2}V^{''''}}{V^{3}}\frac{1}{(1+\frac{V}{2\lambda})^{3}}
=
\frac{U^{2}(\phi)J(\phi)}
{S^{3}(\phi)L^{3}(\phi)}\left(\frac{\phi}{M}\right)^{6},\eea

\begin{figure}[htb]
{\includegraphics[width=7cm, height=5cm] {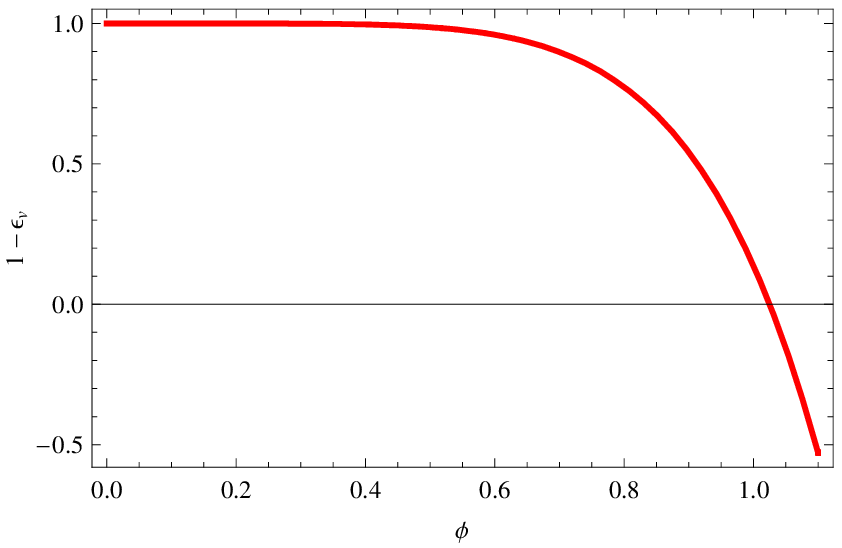},$~~~~$\includegraphics[width=7cm, height=5cm] {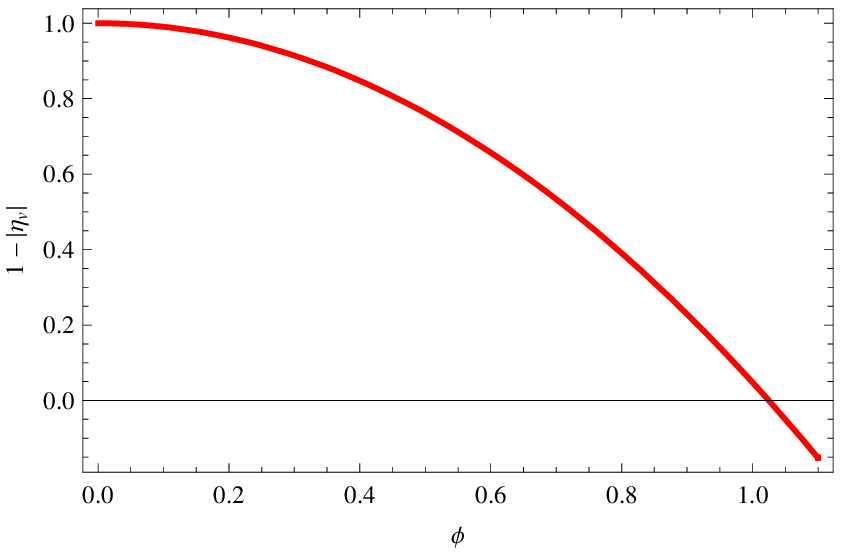}}
\caption{(I) Variation of the 1-$\epsilon_{V}$ vs inflaton field $\phi$ for $C_{4}=-0.68$,  (II)Variation
 of the 1-$|\eta_{V}|$ vs inflaton field $\phi$ for $C_{4}=-0.68$} \label{figVr9}
\end{figure}

Figures (\ref{figVr9})  depict how the first two slow roll parameters vary with the inflaton
field for the allowed range of $D_{4}$ and they give us a clear
picture of the starting point as well as the end of the cosmic
inflation. Nevertheless, Figure (\ref{figVr9}) further reveals
that the $\eta$-problem is smoothened to some extent in brane cosmology. However, we are yet to figure out if there is
any underlying dynamics that may lead to the solution of this
generic feature of SUGRA.

The number of e-foldings are  defined in brane cosmology 
\cite{Maartens} for our model as
 \be\label{noe}
  N=\frac{a(t_{f})}{a(t_{i})}\simeq\frac{8\pi}{M^{2}_{PL}}\int^{\phi_{i}}_{\phi_{f}}\left(\frac{V}{V^{'}}\right)\left(1+\frac{V}{2\lambda}\right)d\phi
 $$$$ \simeq\frac{M^{2}}{U}\left[\frac{1}{2}\left(1+\frac{\alpha}{2}\right)\left(\frac{1}{\phi^{2}_{f}}
-\frac{1}{\phi^{2}_{i}}\right)+\frac{D_{4}}{2M^{4}}(1+\alpha)(\phi^{2}_{i}-\phi^{2}_{f})+
\frac{\alpha
D^{2}_{4}}{12M^{8}}(\phi^{6}_{i}-\phi^{6}_{f})\right]
 \ee
which, in the high energy regime, reduces to
  $ N \simeq \frac{\alpha
M^{2}}{4|U|}\left[\frac{1}{\phi^{2}_{i}}-\frac{1}{\phi^{2}_{f}}\right]$.
Here $\phi_{i}$ and $\phi_{f}$ are the corresponding values of
the inflaton field at the start and end of inflation.

Let us now engage ourselves in analyzing quantum fluctuation in
our model and its observational imprints via primordial spectra
generated from cosmological perturbation \cite{Riazuelo}. In  brane inflation the expressions for
amplitude of the scalar perturbation, tensor perturbation and
tensor to scalar ratio \cite{Maartens}
,\cite{A.A.Sen},\cite{Wands} are given by
 \be\label{scalar}
\Delta^{2}_{s} \simeq
\frac{512\pi}{75M^{6}_{PL}}\left[\frac{V^{3}}
{(V^{'})^{2}}\left[1+\frac{V}{2\lambda}\right]^{3}\right]_{k=aH}
=\frac{M^{2}\alpha\lambda S^{3}(\phi_{\star})L^{3}(\phi_{\star})}{75\pi^{2}U^{2}(\phi_{\star})(\phi_{\star})^{6}},
\ee

\be\label{tensor} \Delta^{2}_{t}  \simeq
\frac{32}{75M^{4}_{PL}}\left[\frac{V\left[1+\frac{V}{2\lambda}\right]}{\left[\sqrt{1+\frac{2V}{\lambda}\left(1+\frac{V}{2\lambda}
\right)}-\frac{2V}{\lambda}\left(1+\frac{V}{2\lambda}\right)\sinh^{-1}\left[\frac{1}{\sqrt{\frac{2V}{\lambda}\left(1+\frac{V}{2\lambda}\right)}}\right]\right]}\right]_{k=aH}
=\frac{\lambda\alpha}{150\pi^{2}M^{4}}\frac{S(\phi_{\star})L(\phi_{\star})}{\tilde{\bar{P}}(\phi_{\star})},
\ee 

\be\label{ratio} r  =
16\frac{\Delta^{2}_{t}}{\Delta^{2}_{s}}\simeq
\frac{8(\phi_{\star})^{6}U^{2}(\phi_{\star})}{M^{6}
 S^{2}(\phi_{\star})L^{2}(\phi_{\star})\tilde{\bar{P}}(\phi_{\star})}.
\ee  

 Here and  throughout the rest of the article $\phi_{\star}$
represents the value of the inflaton field at the horizon
crossing and all the global function defined in eqn(\ref{h341}) is evaluated at the horizon crossing.

\begin{figure}[htb]
{\includegraphics[width=7cm, height=6cm] {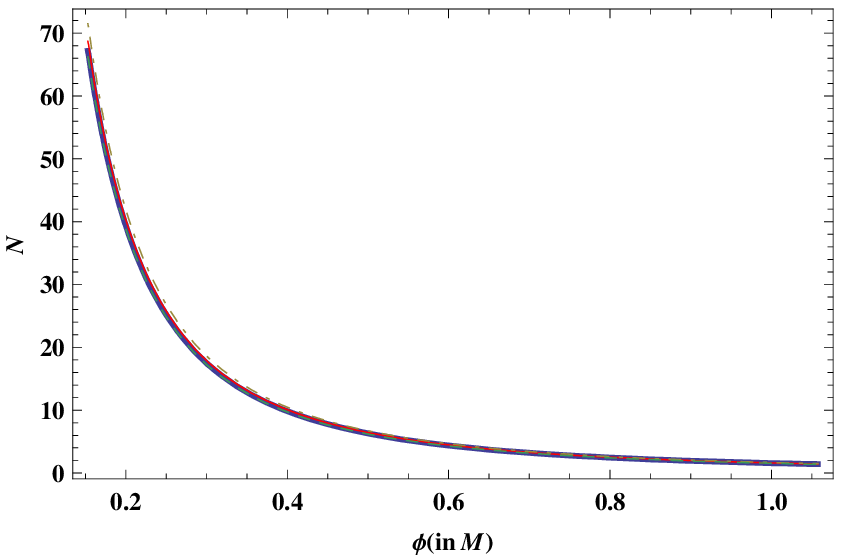}$~~~$\includegraphics[width=7cm, height=6cm] {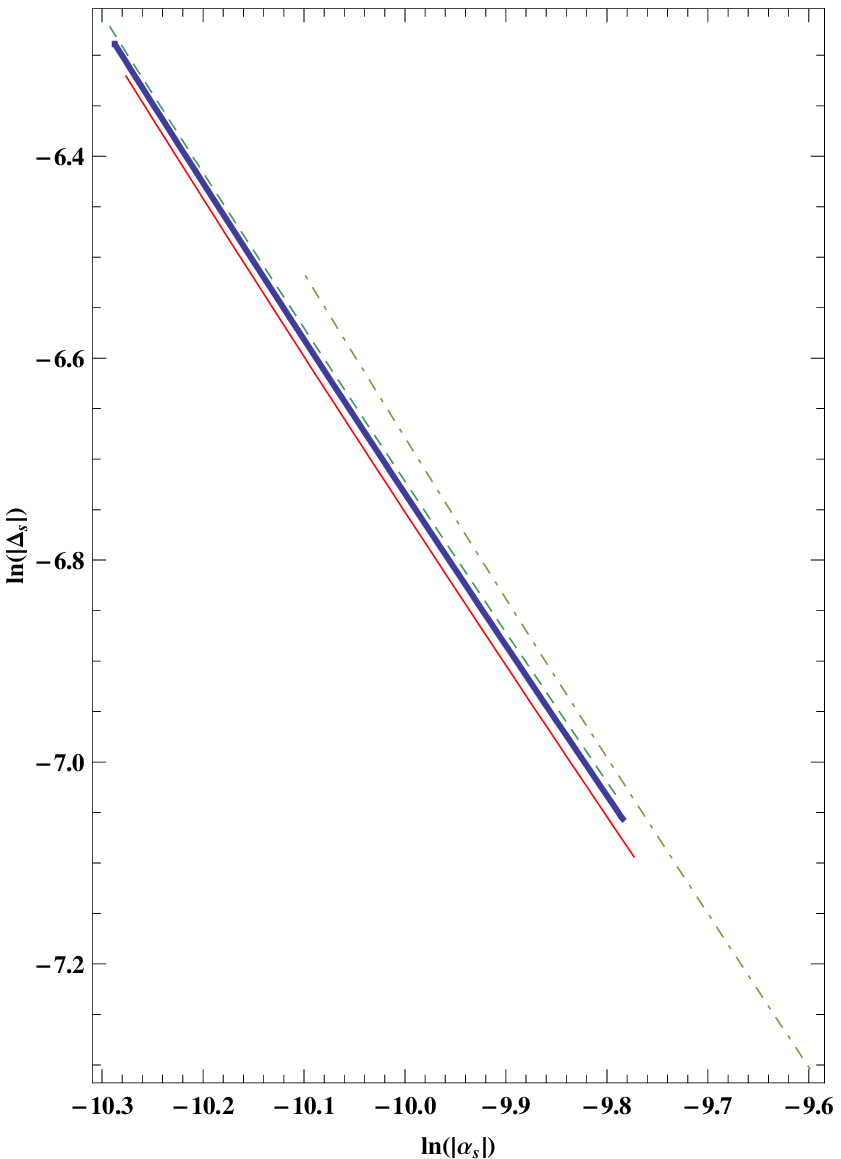}}
\caption{$~$ (I)Variation of the number of e-folding($N$) vs inflation field ($\phi$)(measured in the units of $M$), $~$(II)Variation of the logarithmic scaled amplitude of the scalar fluctuation ($ln({\Delta}_{s})$) vs logarithmic scaled amplitude
of  the running of the spectral index ($\ln(|\alpha_{s}|)$)} \label{figVr241}
\end{figure}

Figure(\ref{figVr241}(I)) represents a graphical behavior of number
of e-folding versus the inflaton field in the high energy limit
for different values of $D_{4}$ and the most satisfactory point in
this context is the number of e-folding lies within the
observational window  $56<N<70$. The end of the inflation leads
to the constraint $\alpha=\frac{2}{(|U|)}(|E|)^{\frac{3}{2}}$ which is required 
for numerical estimations. Here figure(\ref{figVr241}(II)) represents the logarithmically scaled
plots of the physical set of parameter
($\Delta_{s},\alpha_{s}$)for  different values of $D_{4}$. The
plots themselves present good fit with observations.

Further,  the scale dependence of the perturbations, described by
the scalar and tensor spectral indices, as follows
\cite{lythm},\cite{Bassett} \be\label{tilts}
 n_{s}-1=\frac{d(\ln(\Delta^{2}_{s}))}{d
(\ln(k))}\simeq (2\eta^{\star}_{V}-6\epsilon^{\star}_{V})=\frac{2E(\phi_{\star})}
{S(\phi_{\star})L(\phi_{\star})}\left(\frac{\phi_{\star}}{M}\right)^{2}
-\frac{3U(\phi_{\star})T(\phi_{\star})}{S^{2}(\phi_{\star})L^{2}(\phi_{\star})}\left(\frac{\phi_{\star}}{M}\right)^{6},$$$$ ~~~~ n_{t}=\frac{d(\ln
(\Delta^{2}_{t}))}{d(\ln(k))}\simeq -3\epsilon^{\star}_{V}= -\frac{3U^{2}(\phi_{\star})T(\phi_{\star})}{2S^{2}(\phi_{\star})L^{2}(\phi_{\star})}\left(\frac{\phi_{\star}}{M}\right)^{6}.\ee
where $d(\ln(k))=Hdt$. Here one can check that  \cite{santos} the validity of the consistency condition
$r=24\epsilon_{V}=24\epsilon^{\star}_{V} ; ~~
n_{t}=-3\epsilon_{V}\simeq-3\epsilon^{\star}_{V}=-\frac{r}{8}$.

 The expressions for the
running of the scalar and tensor spectral index in this specific 
model with respect to the logarithmic pivot scale at the horizon crossing are given by

 \be\label{r3}
\alpha_{s}=(16\eta\epsilon-18\epsilon^{2}-2\xi)$$$$=\frac{8E(\phi_{\star})U^{2}(\phi_{\star})T(\phi_{\star})}{S^{3}(\phi_{\star})L^{3}(\phi_{\star})}\left(\frac{\phi_{\star}}{M}\right)^{8}
-\frac{2F(\phi)U(\phi_{\star})}
{S^{2}(\phi_{\star})
L^{2}(\phi_{\star})}\left(\frac{\phi_{\star}}{M}\right)^{4}-\frac{9U^{4}(\phi_{\star})T^{2}(\phi_{\star})}{2S^{4}(\phi_{\star})L^{4}(\phi_{\star})}\left(\frac{\phi_{\star}}{M}\right)^{12},
 \ee

\be\label{r3a}
\alpha_{t}=(6\epsilon\eta-9\epsilon^{2})
=\frac{3E(\phi_{\star})U^{2}(\phi_{\star})T(\phi_{\star})}{S^{3}(\phi_{\star})L^{3}(\phi_{\star})}\left(\frac{\phi_{\star}}{M}\right)^{8}-\frac{9U^{4}(\phi_{\star})T^{2}(\phi_{\star})}{4S^{4}(\phi_{\star})L^{4}(\phi_{\star})}\left(\frac{\phi_{\star}}{M}\right)^{12},
 \ee

 One can also calculate the running of the fourth slow roll
parameter as 
$\frac{d\sigma}{d(\ln(k))}=(\epsilon\sigma-2\eta\sigma)$, but
its numerical value  turns out to be too small to be detected even
in near future for which it can treated as consistency condition in  brane.

To estimate five dimensional Planck mass from the observational parameters we use the
relation $\sqrt{8\pi}M=M_{PL}=\frac{M^{3}_{5}}{\sqrt{\lambda}}\sqrt{\frac{3}{4\pi}}$.
and Eq (\ref{scalar})
which leads to \be\label{fiv}
M_{5}=\sqrt[6]{\frac{800\pi^{4}\Delta^{2}_{s}U^{2}(\phi_{\star})}
{\alpha S^{3}(\phi_{\star})L^{3}(\phi_{\star})}}\phi_{\star}.\ee

Finally using the thermodynamic definition of density at the time of reheating 
$\rho(t_{reh})=\frac{\pi^{2}N^{\star}T^{4}_{brh}}{30}$ in the inflaton decay width 
$\Gamma_{total}=3H(T^{breh})=3\sqrt{\frac{\rho(t_{reh})}{3M^{2}}\left[1+\frac{\rho(t_{reh})}{2\lambda}\right]}\simeq
 \frac{\Delta^{6}}{(2\pi)^{3}M^{5}}$ we have estimated the reheating temperature in the braneworld in terms of the five dimensional Planck mass
as
\be\begin{array}{ll}\label{reh}
\displaystyle T^{breh}=\sqrt{\frac{3}{4\pi^{2}}\sqrt{\frac{5}{
N^{*}}}\frac{M^{3}_{5}}{M}}\sqrt[4]{\left[\sqrt{1+\frac{64M^{4}\pi^{2}\Gamma^{2}_{total}}
{9M^{6}_{5}}}-1\right]}\\~~~~~~~\displaystyle =\sqrt[4]{\frac{2250\Delta^{2}_{s}U^{2}(\phi_{\star})\phi^{3}_{\star}}
{N^{*}M^{2}\alpha S^{3}(\phi_{\star})L^{3}(\phi_{\star})}}\sqrt[4]{\left[\sqrt{1+
\frac{2M^{4}\Gamma^{2}_{total}\alpha S^{3}(\phi_{\star})L^{3}(\phi_{\star})}{225\pi^{2}
\Delta^{2}_{s}\phi^{6}_{\star}U^{2}(\phi_{\star})}}-1\right]},\end{array}\ee
where $N^{\star}$ is the effective number of particles incorporating the relativistic degrees of freedom.

\section{parameter estimation}

\subsection{Direct numerical estimation}

   \begin{table}[htb]
   \begin{tabular}{|c|c|c|c|c|c|c|c|c|c|c|c|c|c|c|c|c|}
   \hline $C_{4}$ & $\alpha$  & $\lambda$& $\phi_{f}$ & $\phi_{i}$ & $N$ & $\phi_{\star}$ & $\Delta^{2}_{s}$ & $\Delta^{2}_{t}$ & $n_{s}$ & $n_{t}$ & $r$ & $\alpha_{s}$ &$\alpha_{t}$& $M_{5}$& $T^{breh}$\\
    $\simeq D_{4}$ & & $\times10^{-14}M^{4}$& $M$ & $M$ &  & $M$ & $\times10^{-9}$& $\times10^{-14}$ & & $\times10^{-5}$ &$\times10^{-5}$ & $\times10^{-3}$ &$\times10^{-6}$& $\times10^{-3}M$& $\times10^{-8}M$\\
   \hline
    &  &  &  &0.147 & 70 & 0.158 & 3.126 &  & 0.951 & -4.352 & 2.176 &-0.798 & -2.125 & & \\
   -0.70  &  17.389 & 2.553 &1.017 &0.158& 60 & 0.173 & 1.835 & 6.803 & 0.941 & -7.412 & 3.706 &-1.142 & -4.323 & 11.792 & 3.119  \\
        &  &  &  &0.164 & 56 & 0.180 & 1.440 &  & 0.936 & -9.447 & 4.723 &-1.345 & -5.975& &\\
   \hline
       &  &  &  &0.150 & 70 & 0.161 &2.902 &  & 0.951 & -4.352 &2.176 & -0.798 & -2.125 & & \\
   -0.65 &  16.757 & 2.632 & 1.036 & 0.161& 60 & 0.176 &1.704& 6.317 & 0.941 & -7.412 & 3.706 & -1.142 & -4.323 & 11.865 & 3.133 \\
     &  &  &  &0.167  & 56 & 0.184 &1.327 &  & 0.936 & -9.447 & 4.723 &  -1.345 & -5.975 &  &\\
   \hline
       &  &  &  &0.153 & 70 & 0.165 & 2.679 & & 0.951 & -4.352 &  2.176 & -0.798 &-2.125 & &   \\
   -0.60 &  16.099 & 2.758& 1.057 & 0.165 &60 & 0.180 & 1.573 & 5.831 & 0.941 & -7.412 & 3.706 &-1.142 & -4.323 &11.944 & 3.149 \\
      &  &  &  &0.170 & 56 & 0.187 & 1.234 &  & 0.936 &-9.447 & 4.723&-1.345 &-5.975 & &  \\
   \hline
  \end{tabular}
  \caption{Different observational  parameters related to the cosmological perturbation
  for our  model of inflation including one loop radiative correction }
  \label{tab1}
  \end{table}

Table \ref{tab1} 
represent numerical estimation for different observational
parameters related to the cosmological perturbation as  estimated
 from our model. Here a ``$\times$'' implies ``in units of''. It is worthwhile to point
out to the salient  features of those parameters in the
above  table as obtained from our model.
\begin{itemize}

\item The observable parameters  help us have an
    estimation for the brane tension to be $\lambda\gg (1
    MeV)^{4}$ provided energy scale of the inflation is in the
    vicinity of GUT scale and exactly it is of the order of
    $0.2\times 10^{16}GeV$ which resolves Polonyi problem
    \cite{Coughlan} and Gravitino problem \cite{Chol}.

\item The scalar power spectrum corresponding to different
    best fit values of $D_{4}$ mentioned above is of the order
    of  $5\times 10^{5}$ and it perfectly matches with the
    observational data \cite{wmap07}.

\item The scalar spectral index  for lower values of
    $N\rightarrow 55$ are pretty close to observational window
    $0.948<n_{s}<1$ \cite{wmap07} whereas for higher values of
    $N\rightarrow 70$  this lies well within the window.
    Thus this small observational window  reveals that $N \approx
    70$ is more favored in brane cosmology compared to its lower
    values.

\item Though the tensor to scalar ratio as estimated from our
    model is well within its upper bound fixed by WMAP7
    \cite{wmap07}  ($r<0.45$ at $95 \%$ C.L.), thereby facing no contradiction
    with observations, its value is even small to be detected
    in WMAP \cite{wmap07}  or the forthcoming Planck
    \cite{planck}. For more discussion see \cite{grav}.

\item For our model running of the scalar spectral index
    $\alpha_{s} \sim -10^{-3}$ which is quite consistent with
    WMAP3 \cite{wmap3}. Also,  the running of the tensor
    spectral index $\alpha_{t} \sim-6\times 10^{-6}$ may serve
    as an additional observable parameter to be investigated
    further.

\item Five dimensional Planck mass turns out to be $M_{5}\sim (11.792-11.944)\times 10^{-3}M$
      which is the prime input for the estimation of brane reheating temperature as shown in 
      eqn(\ref{reh}). For our model it is estimated as $T^{breh}\sim(3.119-3.149)\times 10^{-8}M$
      and clearly depicts the deviation from standard cosmology.

\end{itemize}

\subsection{Data analysis with CAMB}

In this context we shall make use of the cosmological code CAMB\cite{camb} in order
 to confront our results directly with observation.
To operate CAMB, the values of the initial parameters
associated with inflation are taken from the Table\ref{tab1} for $D_{4}=-0.60$. Additionally WMAP7 dataset in 
$\Lambda$CDM background has been used in CAMB to obtain CMB angular power spectrum at the pivot scale $k_0=0.002~{\rm Mpc}^{-1}$.
Table \ref{tab2} and table\ref{tab3} shows input from the WMAP7 dataset and 
the output obtained from CAMB respectively. 



\begin{table}
\parbox{.46\linewidth}{
\centering
\begin{tabular}{|c|c|c|c|c|c|c|c|c|c|}
\hline $H_0$ & $\tau_{Reion}$ &$\Omega_b h^2$& $\Omega_c h^2
$& $T_{CMB}$
 \\
km/sec/MPc& & && K\\
 \hline
71.0&0.09&0.0226&0.1119&2.725\\
\hline
\end{tabular}
\caption{Input parameters in CAMB}\label{tab2}
}
\hfill
\parbox{.46\linewidth}{
\centering
\begin{tabular}{|c|c|c|c|c|c|c|c|c|c|}
\hline $t_0$ & $z_{Reion}$ &$\Omega_m$&$\Omega_{\Lambda}$&$\Omega_k$&$\eta_{Rec}$& $\eta_0$
 \\
Gyr& & && &Mpc & Mpc\\
 \hline
13.707&10.704&0.2670&0.7329&0.0&285.10&14345.1\\
\hline
\end{tabular}
\caption{Output parameters from CAMB}\label{tab3}
}
\end{table}

\begin{figure}[htb]
\centerline{\includegraphics[width=6.1cm, height=5cm]{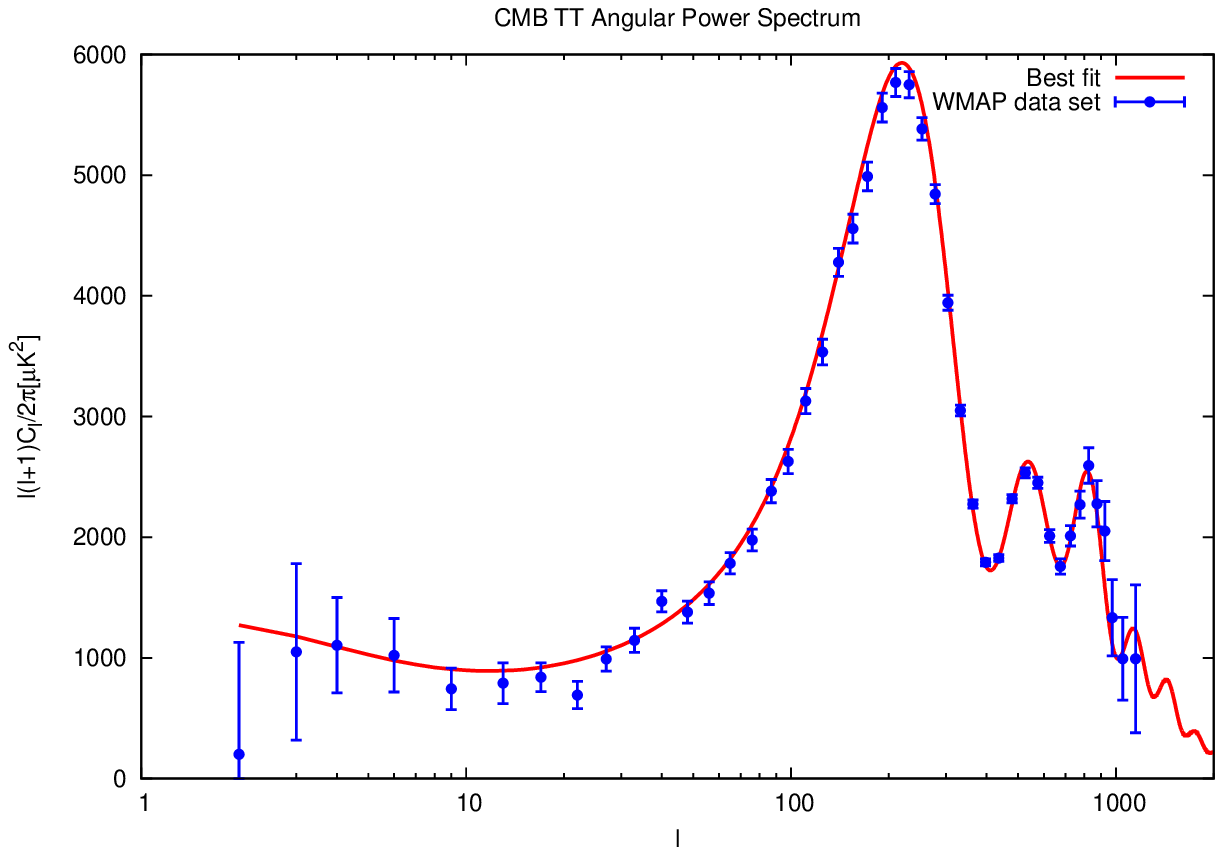} \includegraphics[width=6.1cm, height=5cm]{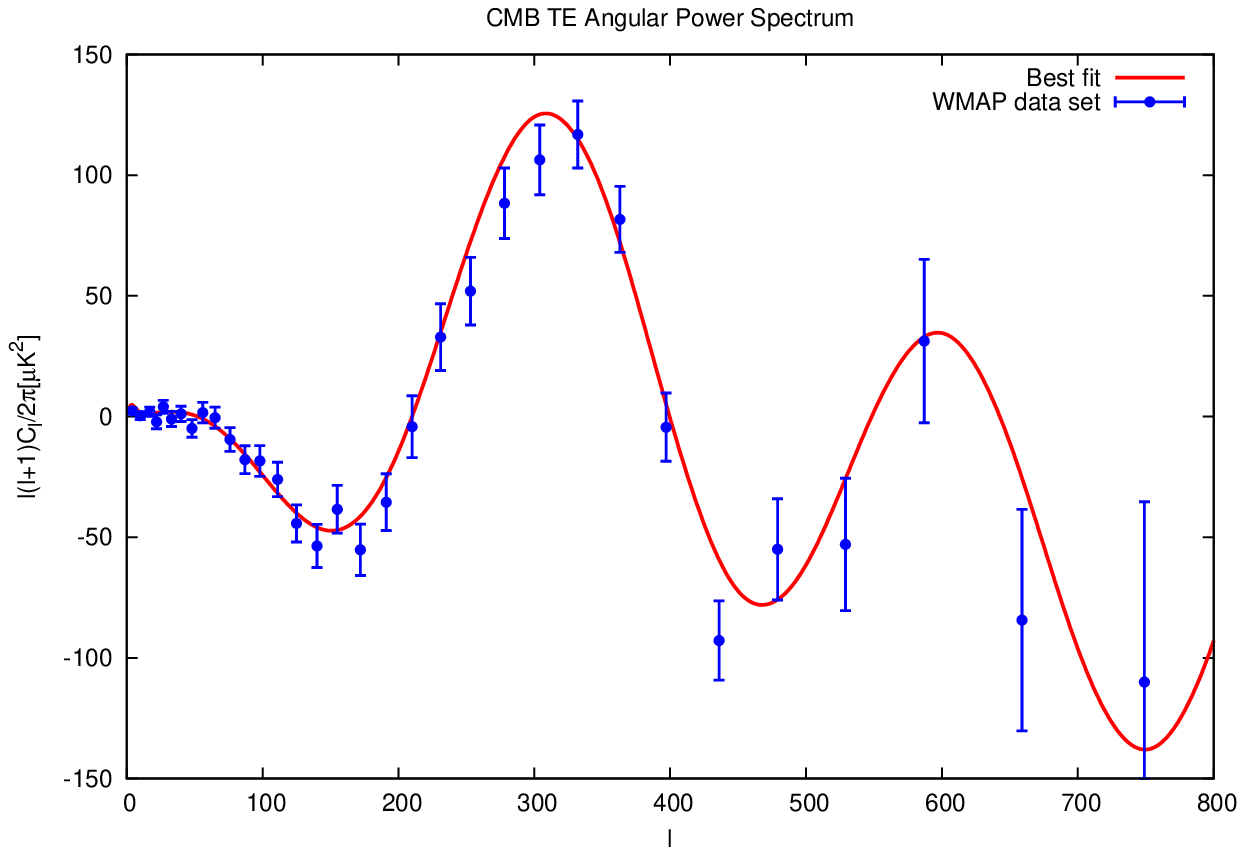} \includegraphics[width=6.1cm, height=5cm]{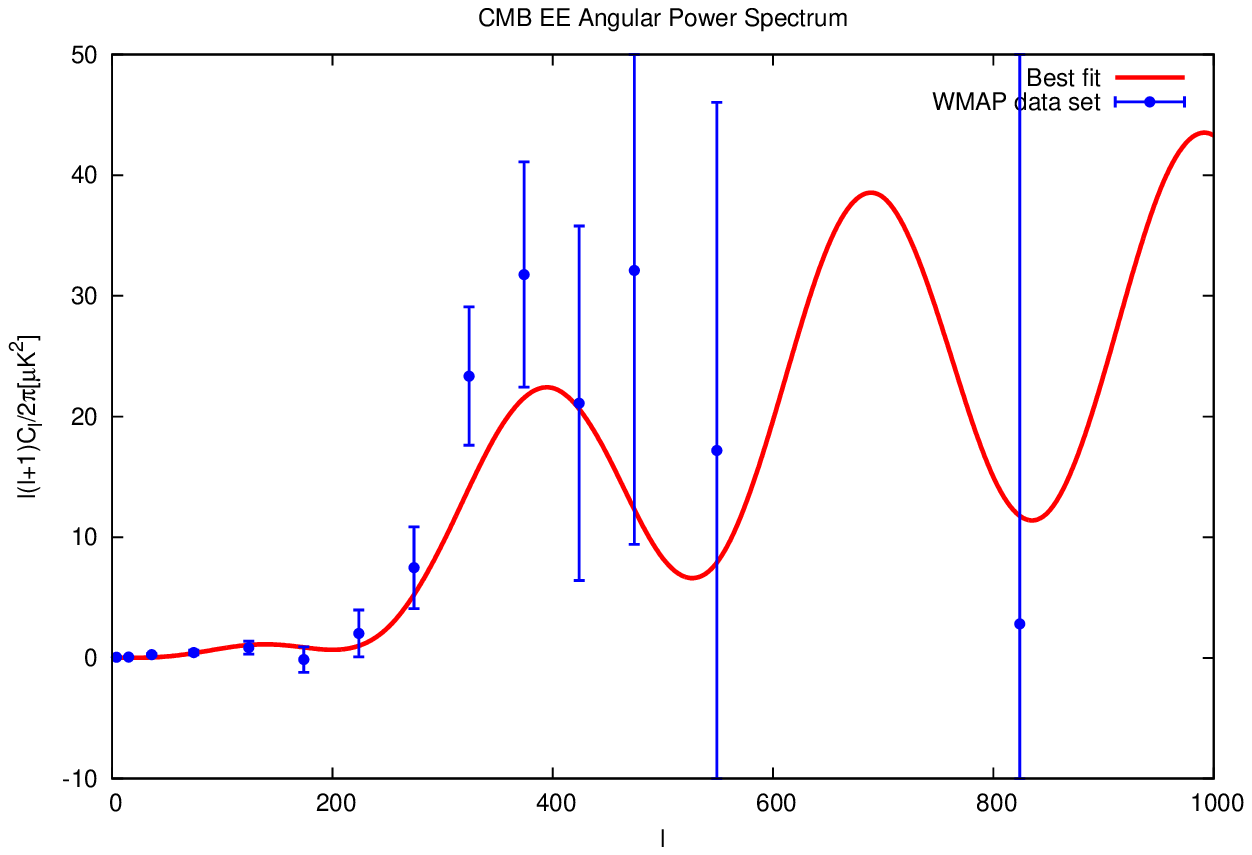}}
 \caption{ Variation of CMB angular power spectrum (a)$C_l^{TT}$, (b)$C_l^{TE}$ and (c)$C_l^{EE}$
 for best fit  and WMAP seven years data with the multipoles $l$ for scalar modes}\label{tt}
\end{figure}
\begin{figure}[htb]
 \centerline{\includegraphics[width=7cm, height=5cm]{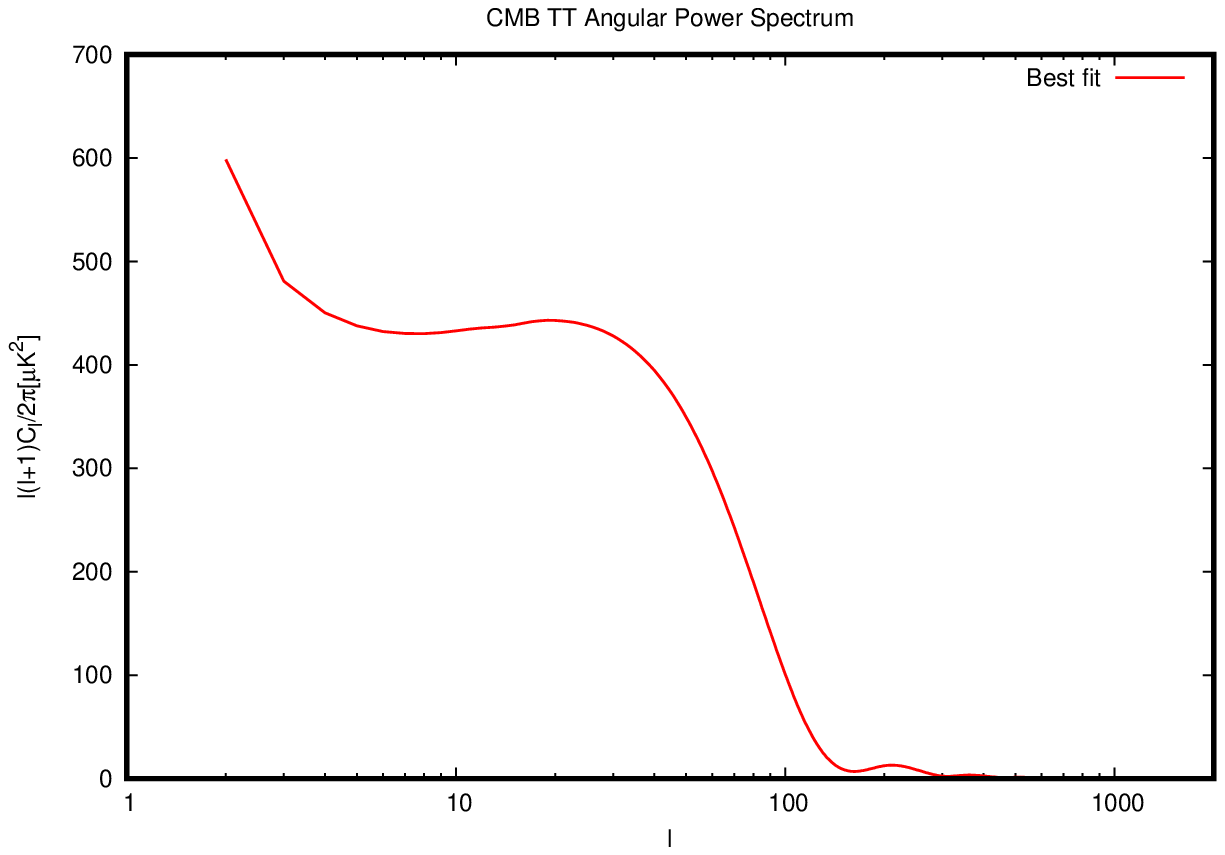}$~~~~~~$\includegraphics[width=7cm, height=5cm]{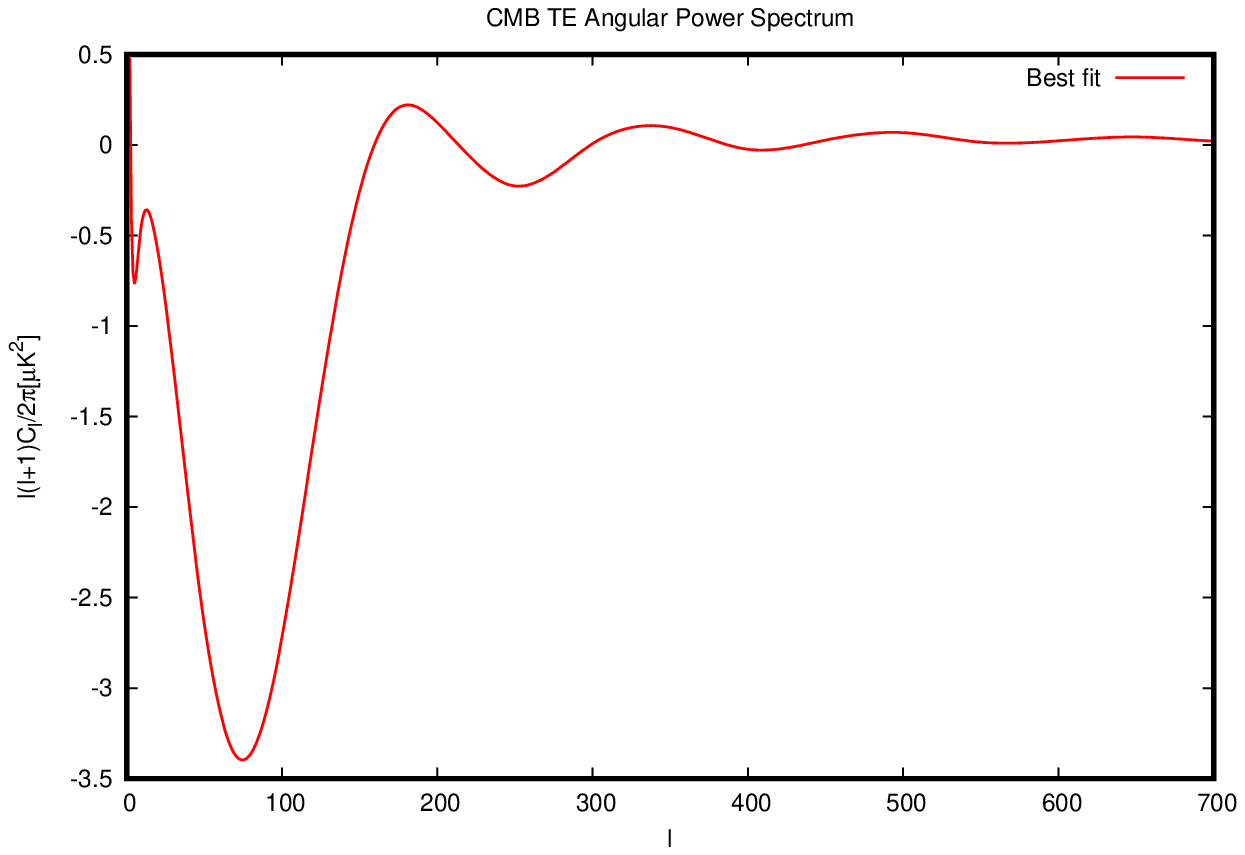}}
 \centerline{\includegraphics[width=7cm, height=5cm]{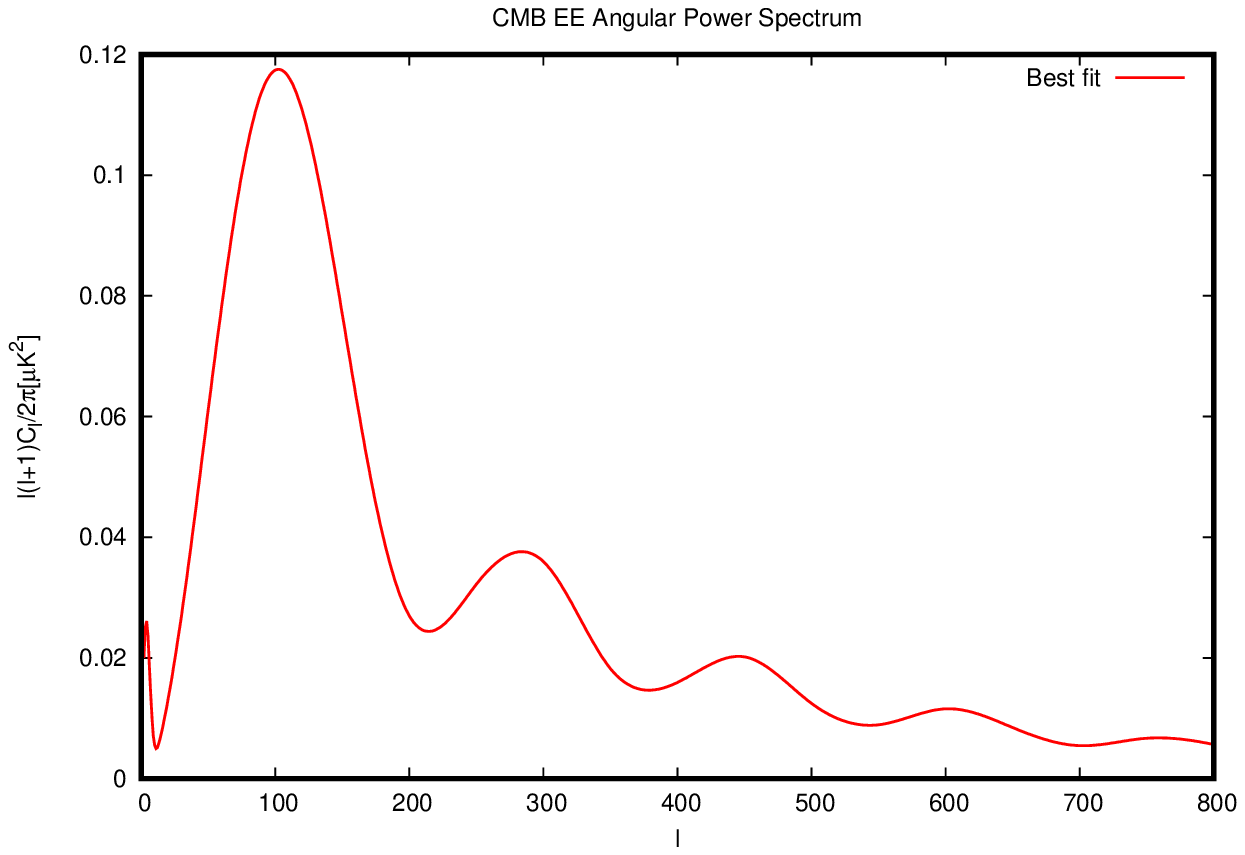}$~~~~~~$\includegraphics[width=7cm, height=5cm]{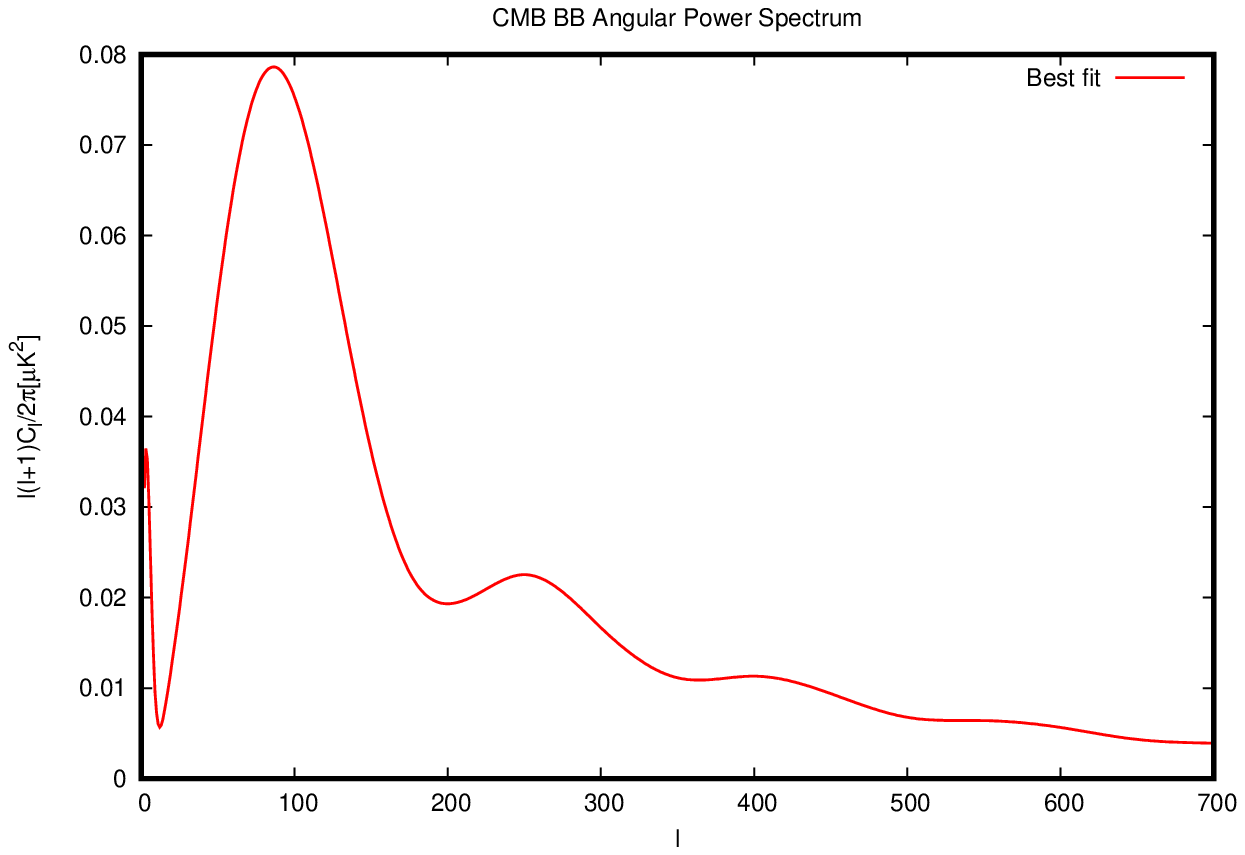}}
  \caption{ Variation of CMB angular power spectrum (a)$C_l^{TT}$, (b)$C_l^{TE}$,
  (c)$C_l^{EE}$ and (d)$C_l^{BB}$ with the multipoles $l$ for tensor mode}\label{ttt}
\end{figure}

The curvature perturbation is generated due to the fluctuations in
the {\it inflaton} and at the end of inflation it makes horizon
re-entry creating matter density fluctuations, which is the origin of the structure formation in
Universe. In Fig.\ref{tt}(a)-Fig\ref{tt}(c) we confront CAMB output of CMB angular power spectrum $C_l^{TT}$, $C_l^{TE}$ and $C_l^{EE}$
 for best fit with WMAP seven years data for the scalar mode.
 From Fig.\ref{tt}(a) we see that the Sachs-Wolfe plateau \cite{sachs} obtained from
  our model is almost flat confirming a nearly scale invariant spectrum. For larger value of the multipole $l$,
 CMB anisotropy spectrum is dominated by the
Baryon Acoustic Oscillations (BAO) \cite{mukhanovb} giving rise to
several ups and downs in the spectrum. Also the peak positions
are sensitive on the dark energy and other forms of
the matter. In Fig.\ref{tt}(a) the first and most prominent peak
arises at $l=221$ at a height of $5818{\rm \mu K^2}$ followed by
two equal height peaks at $l=529$ and $l=822$. This is in good
 agreement with WMAP7 data for $\Lambda$CDM
background apart from the two
outliers at $l=21$ and $l=42$. The gravitational waves generated during inflation also remain
constant on {\it super Hubble} scales having small amplitudes which die off very rapidly
due to smaller wavelength than horizon. So the small scale modes have no
impact in the CMB anisotropy spectrum only the large scale modes
have little contribution and this is obvious from Fig.\ref{ttt}(a)-Fig.\ref{ttt}(d) where 
we have plotted the CAMB output of CMB angular power spectrum $C_l^{TT}$, $C_l^{TE}$, $C_l^{EE}$ and $C_l^{BB}$
 for best fit with WMAP7 data for the tensor mode. Thus, from the entire data analysis with CAMB, it turns out that our model
 confronts extremely well with WMAP7 dataset and leads
 to constrain the best fit value of the parameter $D_{4}$ at $-0.60$.


\section{\bf Dynamical signature of the model}

Let us now engage ourselves in
finding out the dynamical signature of the model from the first
principle. Precisely, we are interested  to obtain a solution of the modified
Friedman equation and Klein-Gordon equation in brane
cosmology with our proposed model. Under slow-roll approximations
 the inflaton field as a function of cosmic time can be expressed as \be
\label{sol1}\phi(t)=\frac{M^{2}}{\sqrt{2D_{4}}}\sqrt{\left[\tilde{\bar{\Phi}}(f)-\bar{G}t\right]}
\sqrt{\left[1-\sqrt{1+\frac{4D_{4}}{M^{4}\left[\tilde{\bar{\Phi}}(f)-\bar{G}t\right]^{2}}}\right]},\ee
where $ \bar{G}=\frac{2U\sqrt{2\lambda}}{\sqrt{3}M^{3}},~~
\tilde{\bar{\Phi}}(f)=\frac{1}{\phi^{2}_{f}}\left(\frac{D_{4}\phi^{4}_{fe}}{M^{4}}-1\right)+\bar{G}t_{f}$.
It may be noted that in the high energy limit, the above
equation(\ref{sol1}) reduces to a much  tractable form
$\phi(t)=\phi_{f}\left[1+\frac{2U\phi^{2}_{f}}{M^{3}}\sqrt{\frac{2\lambda}{3}}(t-t_{f})\right]^{-\frac{1}{2}}$.

\begin{figure}[htb]
{\includegraphics[width=7cm, height=5cm] {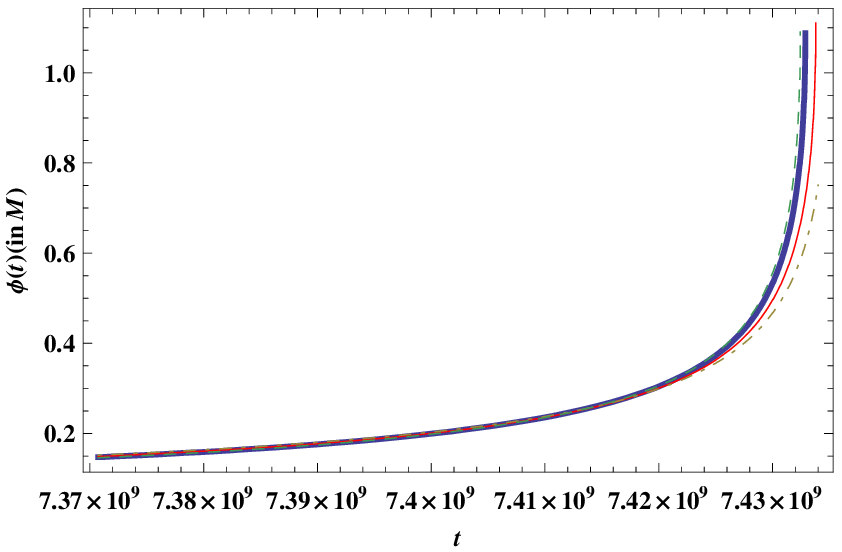}$~~~~$\includegraphics[width=7cm, height=5cm] {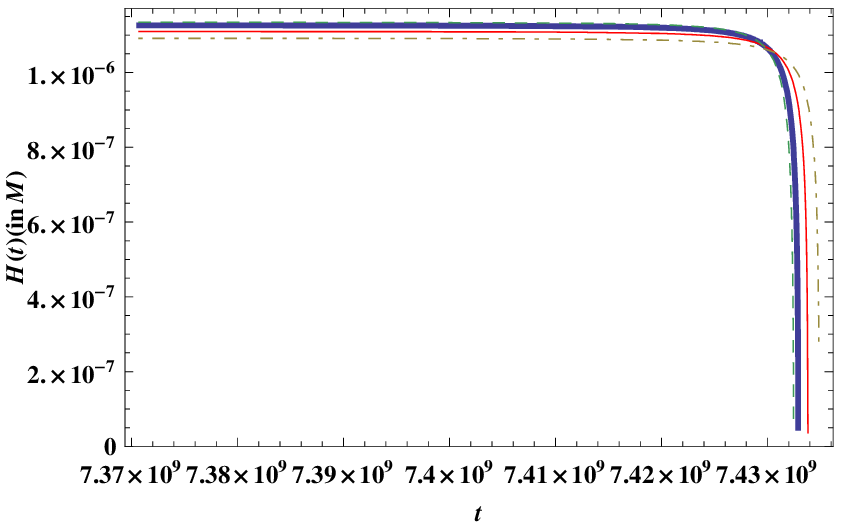}}
\caption{ (I)$~$Variation of the inflaton field ($\phi$) with time(t), 
(II)$~$ Variation of the Hubble parameter
 ($H(t)$) with time($t$)} \label{figVr230}
\end{figure}

 Figure (\ref{figVr230}(I)) shows the evolution of the inflaton
field under high energy approximation which shows a smooth
increasing behavior of the inflaton field with respect to the
inflationary time scale where the span of  the scale are within
the window $t_{i}<t<t_{f}$. In figure (\ref{figVr230}(II)) the
 evolution of the Hubble parameter shows deviations from the
 de-Sitter   as given  by the bending of the
 plots towards the end of inflation which leads to physically more realistic scenario so as to fit with observational
data as demonstrated earlier.

Substituting equation(\ref{sol1}) in the modified Friedman
equation in brane for our model we obtain
\be\label{hub1}H(t)=\sqrt{\frac{\lambda}{6}}\frac{\alpha}{M}\left[2+\frac{M^{4}}
{2D_{4}}\left[\tilde{\bar{\Phi}}(f)-\bar{G}t\right]^{2}\left(1-\sqrt{1+\frac{4D_{4}}{M^{4}\left[\tilde{\bar{\Phi}}(f)-\bar{G}t\right]^{2}}}\right)\right]\ee
which shows the time evolution as well as the susceptance of
Hubble parameter in the context of brane.

Consequently, the solution of the modified Friedman equation,
after rearranging terms,  gives rise to the scale factor as
follows
\be\label{fr1}a(t)=a(t_{f})\exp\left[\sqrt{\frac{\lambda}{6}}\frac{\alpha}{M}\left[2(t-t_{f})+\tilde{A}(t-t_{f})+\frac{\tilde{B}}{3}(t^{3}-t^{3}_{f})-\frac{\tilde{C}}{2}(t^{2}-t^{2}_{f})
-\tilde{I}(t)\right]\right]\ee where $ \tilde{I}(t)=
\int^{t}_{t_{f}}dt\sqrt{\left[(\tilde{A}+\tilde{B}t^{2}-\tilde{C}t+1)^{2}-1\right]},~
\tilde{A}=\frac{M^{4}\tilde{\bar{\Phi}}(f)}{2D_{4}}, \tilde{B}=\frac{\bar{G}^{2}M^{4}}{2D_{4}},~
\tilde{C}=\frac{\tilde{\bar{\Phi}}(f)\bar{G}M^{4}}{D_{4}}$. Thus the scale factor can
be obtained analytically except for the integrand $\tilde I(t)$, and it
readily shows the deviation from the standard de Sitter model.
However, the above form of the scale factor (\ref{fr1}) is more or
less sufficient to study the dynamical behavior, as represented in
Figure(\ref{figVr230}(II)). As a matter of fact, the leading order
contribution from Hubble parameter and the scale factor are indeed
closed to de Sitter with the parameters involving brane cosmology.



\section{\bf Analysis of the energy scale of brane inflation}

Let us now estimate the typical scale of inflation in brane
cosmology with the potential of our consideration.  For this we
shall make use of two initial conditions, namely, initial time
    $t_{i}=0.737\times10^{10}M^{-1}$ and
    $a(t_{i})=0.369\times10^{-1}M^{-1}$.
Consequently, for $N=70$ we have
  $a(t_{f})=0.929\times10^{11}M^{-1}$. Now taking leading order contribution from  Eq
(\ref{fr1}) the time corresponding to the horizon exit and re-entry can be obtained as
\be\label{tstar}
t_{\star}=t_{f}+\frac{1}{\bar{G}}\left[\tilde{\bar{\Phi}}(f)-\frac{\left[1\pm\sqrt{1-8D_{4}\left[(\phi_{\star})^{2}+2M^{4}\right]}\right]}{M^{4}}\right],\ee

 with $t_{f}=t_{i}+\frac{NM}{\alpha}\sqrt{\frac{6}{\lambda}}$. Using Eq (\ref{tstar}), Eq (\ref{sol1}) and Eq (\ref{second}) 
energy scale of brane inflation can be expressed as
\be\label{highana}\Delta \approx \sqrt[4]{\left[\frac{2E\lambda\phi^{2}_{f}}{|\eta_{V}|M^{2}\left[1+
\frac{2U\phi^{2}_{f}}{M^{3}}\sqrt{\frac{2\lambda}{3}}(t-t_{f})\right]}\right]}.\ee

\begin{figure}[htb]
{\centerline{\includegraphics[width=7cm, height=5cm] {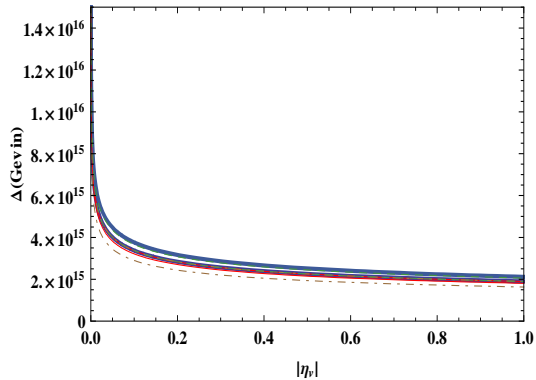}}}
\caption{Variation of the energy scale of inflation ($\Delta$) vs $|\eta_{V}|$ including two roots of the horizon crossing time for the best fit model} \label{figVr235}
\end{figure}

Figure (\ref{figVr235}) shows  the energy scale of inflation
($\Delta$) versus the magnitude of the second slow roll
parameter($|\eta_{v}|$) for different values of the constant
$D_{4}$ including two feasible roots of horizon crossing. From the
figure it is obvious that for two feasible roots of time
corresponding to the horizon crossing an allowed region with
finite band-width appears for our proposed model. The above figure further reveals that the typical energy scale of
brane inflation with our proposed model is $\Delta\simeq
2\times10^{15}GeV$ which is supported from cosmological as well as
particle physics frameworks.


\section{Summary and outlook}

In this article we have proposed a model of inflation in brane
cosmology. We have  demonstrated how we can construct an effective
4D inflationary potential starting from $N=2, D=5$ supergravity in
the bulk which leads to an effective $N=1, D=4$ supergravity in the brane. After 
that we have engaged ourselves in analyzing radiative
corrections of the tree level potential and the effective potential
calculated from one loop correction has then been employed in
estimating the observable parameters, both analytically and
numerically, leading to more precise estimation of the quantities
 and confronting them with WMAP7 dataset
using the publicly available code CAMB, which
reveals consistency of our model with latest observations. 
The increase in precision  level is worth analyzing considering
the advent of more and more sophisticated techniques, both in WMAP
\cite{wmap07} and in forthcoming Planck \cite{planck} data.

We
have also  solved the modified Friedmann equations on
the brane leading to an analytical expression for the scale factor
during inflation.
Finally we have estimated the typical energy scale of brane
inflation with the potential of our consideration and found it to
be consistent with cosmological as well as particle physics
frameworks. This model thus leads to an inflationary scenario
in the framework of supergravity inspired brane cosmology.

A detailed survey of thermal history of the universe via reheating,
 baryogenesis, leptogenesis with the loop corrected potential and 
gravitino phenomenology remains as an open issue, which may even
 provide interesting signatures of brane
inflation. A detailed analysis on these aspects have been reported as a separate paper \cite{reheat}.


\section*{Acknowledgments}

SC thanks S. Ghosh, R. Gopakumar, A. Mukhopadhyay and B. K. Pal
for illuminating discussions and Council of Scientific and
Industrial Research, India for financial support through Junior
Research Fellowship (Grant No. 09/093(0132)/2010). SP is supported
by Alexander von Humboldt Foundation, Germany through the project
``Cosmology with Branes and Higher Dimensions'' and is partially
supported by the SFB-Tansregio TR33 ``The Dark Universe''
(Deutsche Forschungsgemeinschaft) and the European Union 7th
network program ``Unification in the LHC era''
(PITN-GA-2009-237920). Special thanks to Hans Peter Nilles for a
careful reading of the manuscript and for his valuable suggestions
towards the improvement of the article.

\section{appendix}

For systematic development of the formalism, let us  demonstrate
briefly how one can construct the effective 4D inflationary
potential of our consideration starting from $N=2, D=5$ SUGRA in
the bulk which leads to an effective $N=1, D=4$ SUGRA in the brane. As
mentioned, we consider the bulk to be five dimensional where the
fifth dimension is compactified on the orbifold $S^{1}/Z_{2}$ of
comoving radius R. The system is described by the following action \cite{A.Riotto},
\cite{kanti} \be\label{as1} S=\frac{1}{2}\int d^{4}x\int^{+\pi
R}_{-\pi
R}dy\sqrt{g_{5}}\left[M^{3}_{5}\left(R_{(5)}-2\Lambda_{5}\right)
+\textit{\textsl{L}}_{bulk}+\sum_{i}\delta(y-y_{i})\textit{\textsl{L}}_{4i}\right].\ee
Here the sum includes the walls at the orbifold points
$y_{i}=(0,\pi R)$ and 5-dimensional coordinates
$x^{m}=(x^{\alpha},y)$, where $y$ parameterizes the extra
dimension compactified on the closed interval $[-\pi R,+\pi R]$
and $Z_{2}$ symmetry is imposed. For $N=2, D=5$
supergravity in the bulk Eq (\ref{as1}) can be written as
\be\label{su1}
 S=\frac{1}{2}\int d^{4}x\int^{+\pi
R}_{-\pi R}dy\sqrt{g_{5}}\left[M^{3}_{5}\left(R_{(5)}-2\Lambda_{5}\right)+L^{(5)}_{SUGRA}+\sum_{i}
\delta(y-y_{i})L_{4i}\right],\ee which is a generalization of the
scenario described in \cite{A.Riotto}. Written explicitly, the
contribution from bulk SUGRA in the action is given by
\cite{Georg}
\be\label{sug2}e^{-1}_{(5)}L^{(5)}_{SUGRA}=-\frac{M^{3}_{5}R^{(5)}}{2}+\frac{i}{2}\bar{\Psi}_{i\tilde{m}}
\Gamma^{\tilde{m}\tilde{n}\tilde{q}}\nabla_{\tilde{n}}
\Psi^{i}_{\tilde{q}}-{S}_{IJ}F^{I}_{\tilde{m}\tilde{n}}F^{I\tilde{m}\tilde{n}}-\frac{1}{2}
g_{\alpha\beta}(D_{\tilde{m}}\phi^{\mu})(D^{\tilde{m}}\phi^{\nu})
$$$$ + {\rm Fermionic} + {\rm Chern-Simons},\ee Including the
contribution from the radion fields $\chi = -\psi_{5}^{2}$ and $T
= \frac{1}{\sqrt{2}} \left(e_{5}^{\dot 5} - i \sqrt{\frac{2}{3}}
A_{5}^{0} \right)$ the effective brane SUGRA counterpart turns out
to be \be\label{lsug4}
\delta(y)L_{4}=-e_{(5)}\Delta(y)\left[(\partial_{\alpha}\phi)^{\dagger}
(\partial^{\alpha}\phi)+i\bar{\chi}\bar{\sigma}^{\alpha}D_{\alpha}\chi\right].\ee
Here $\Delta(y)=e^{5}_{\dot{5}}\delta(y)$ is the modified Dirac
delta function which satisfies the normalization
conditions $\int^{+\pi R}_{-\pi R}dy
~e^{5}_{\dot{5}}\Delta(y)=1, ~~ \int^{+\pi R}_{-\pi
R}dy~e^{5}_{\dot{5}}= {\cal L}$ where ${\cal L}$ is the 5
dimensional volume. The Chern-Simons terms can be gauged away
assuming cubic constraints \cite{Georg, Lahanas} and $Z_2$
symmetry.  It is useful to define the five
dimensional generalized $K\ddot{a}hler$ function($G$) in this context as
\cite{Georg, Lahanas}
$G=-3\ln\left(\frac{T+T^{\dagger}}{\sqrt{2}}\right)+\delta(y)\frac{\sqrt{2}}{T+T^{\dagger}}K(\phi,\phi^{\dagger})$, 
 which precisely represents  interaction of the radion with gauge
fields. Including the kinetic term of the five dimensional field
$\phi$ the singular terms  measured from the modified Dirac delta
function can be rearranged into a perfect square thereby leading
to the following expression for the action \be\label{modsug}
S\supset\frac{1}{2}\int d^{4}x\int^{+\pi R}_{-\pi
R}dy\sqrt{g_{5}}e_{(4)}e^{5}_{\dot{5}}\left[g^{\alpha\beta}G_{m}^{n}(\partial_{\alpha}\phi^{m})^{\dagger}(\partial_{\beta}\phi_{n})
+\frac{1}{g_{55}}\left(\partial_{5}\phi-\sqrt{H(G)}\Delta(y)\right)^{2}\right],\ee
where
$H(G)=\exp\left(\frac{G}{M^{2}}\right)\left[\left(\frac{\partial
W}{\partial \phi_{m}}+\frac{\partial G}{\partial
\phi_{m}}\frac{W}{M^{2}}\right)^{\dagger}(G_{m}^{n})^{-1}\left(\frac{\partial
W}{\partial \phi^{n}}+\frac{\partial G}{\partial
\phi^{n}}\frac{W}{M^{2}}\right)-3\frac{|W|^{2}}{M^{2}}\right]$.
It is worthwhile to mention that from eqn(\ref{modsug}) we can compute energy momentum tensor for $N=2,D=5$ SUGRA can be expressed as
\be\label{em1}T_{\alpha\beta}=G^{n}_{m}(\partial_{\alpha}\phi^{m})^{\dagger}(\partial_{\beta}\phi_{n})
-g_{\alpha\beta}\left[g^{\rho\sigma}(\partial_{\rho}\phi^{m})^{\dagger}(\partial_{\sigma}\phi_{n})G^{n}_{m}
+g^{55}(\partial_{5}\phi-\sqrt{H(G)}\Delta(y))^{2}\right],\ee
\be\label{em2}T_{55}=\frac{1}{2}(\partial_{5}\phi-\sqrt{H(G)}\Delta(y))^{2}
-\frac{1}{2}g_{55}g^{\rho\sigma}G^{n}_{m}(\partial_{\rho}\phi^{m})^{\dagger}(\partial_{\sigma}\phi_{n}).\ee
On the other hand by varying the action written in eqn(\ref{modsug}) with respect to the scalar field $\phi$ 
the equation of motion for $N=2,D=5$ SUGRA can be expressed as
\be\label{eqmos}\partial_{5}\left[\frac{e^{5}_{\dot{5}}\sqrt{g_{5}}}{g_{55}}(\partial_{5}\phi-\sqrt{H(G)}\Delta(y))\right]
+\sum_{n} e^{5}_{\dot{5}}\left\{\partial_{\beta}\left[\sqrt{g_{5}}g^{\alpha\beta}G^{n}_{m}(\partial_{\alpha}\phi^{m})\right]
-\frac{\sqrt{g_{5}}}{g_{55}}\Delta(y)\partial_{n}(\sqrt{H(G)})(\partial_{5}\phi-\sqrt{H(G)}\Delta(y))\right\}=0 .\ee
Further, imposing $Z_{2}$ symmetry to $\phi$ via
$\phi(0)=\phi(\pi
R)=0$ and compactifying around a circle $(S^1)$
$\partial_{5}\phi=\sqrt{H(G)}\left(\Delta(y)-\frac{1}{2\pi
R}\right)$ we get,

\be\label{tout}
S=\frac{1}{2}\int d^{4}x\int^{+\pi R}_{-\pi
R}dy\sqrt{g_{5}}\left[M^{3}_{5}\left(R_{(5)}-2\Lambda_{5}\right)
+e_{(4)}e^{5}_{\dot{5}}\left\{g^{\alpha\beta}G_{m}^{n}(\partial_{\alpha}\phi^{m})^{\dagger}(\partial_{\beta}\phi_{n})
-g^{55}\frac{H(G)}{4\pi^{2}R^{2}}\right\}\right].\ee
To discuss elaborately the dimensional reduction technique in the regularized fashion here we have to mention the metric structure in $D=5$
in conformal form is given by,
\be\label{metrix}ds^{2}_{5}=e^{2A(y)}\left(ds^{2}_{4}+R^{2}\beta^{2}dy^{2}\right),\ee
where the D=4 metric $ds^{2}_{4}=g^{\alpha\beta}dx_{\alpha}dx_{\beta}$ is the well known FLRW metric.
 The numerical constant $\beta$ has been introduced just for convenience and physically determines the slope of the 
warp factor $e^{2A(y)} $. Consequently we can express the solution of D=5 Einstein eqns.
explicitly in terms of $\beta$ when the warp factor can be expressed as 
\be\label{warp}e^{2A(y)}=\frac{b^{2}_{0}}{R^{2}\left(e^{\beta y}+\frac{\Lambda_{5}b^{4}_{0}}{24R^{2}}e^{-\beta y}\right)},\ee
 where $b_{0}$ is a constant having dimension of length. Now to trace out all the significant contribution from the fifth dimension using dimensional reduction technique
here we use method of separation of variable $\phi^{m}=\phi(x^{\mu},y)=\phi(x^{\mu})\chi(y)$ which leads to,

\be\begin{array}{lllll}\label{ast8}\displaystyle S=\frac{1}{2}\int d^{4}x\sqrt{g_{4}}\int^{+\pi R}_{-\pi
R}dy \left\{\beta M^{3}_{5}R e^{3A(y)}\left[R_{(4)}-\frac{12}{\beta^{2}R^{2}}\left(\frac{dA(y)}{dy}\right)^{2}-\frac{8}{\beta^{2}R^{2}}
\left(\frac{d^{2}A(y)}{dy^{2}}\right)-2\Lambda_{5}e^{2A(y)}\right]\right.\\ \left.\displaystyle~~~~~~~~+\frac{e_{4}}{b_{0}}\Delta(y)(\partial_{\alpha}
\phi^{\mu})^{\dag}(\partial^{\alpha}\phi_{\mu})\left(\frac{\partial^{2}
K(\phi,\phi^{\dagger})}{\partial\phi^{\dagger}_{\mu}\partial\phi^{\nu}}\right)+C(T,T^{\dagger})\frac{e_{4}}{b_{0}}\frac{\Delta(y)}{4\pi^{2}R^{2}}
e^{\frac{K(\phi,\phi^{\dagger})}{M^{2}}}\left[\left(\frac{\partial W}{\partial
\phi_{\alpha}}+\left(\frac{\partial K(\phi,\phi^{\dagger})}{\partial
\phi_{\alpha}}\right)\frac{W}{M^{2}}\right)^{\dag}\right.\right.
\\ \left.\left.~~~~~~~~~~~~~~~~~~~~~~~~~~~~~~~~~~~~~~~~~~~~~~~~~~~~~~~~~~~~~~~
\displaystyle\times\left(\frac{\partial^{2}K(\phi,\phi^{\dagger})}
{\partial\phi^{\alpha}\partial\phi^{\dagger}_{\beta}}\right)^{-1}
\left(\frac{\partial W}{\partial
\phi^{\beta}}+\left(\frac{\partial K(\phi,\phi^{\dagger})}{\partial
\phi^{\beta}}\right)\frac{W}{M^{2}}\right)-3\frac{|W|^{2}}{M^{2}}\right]\right\},
\\ \displaystyle =\frac{1}{2}\int d^{4}x\sqrt{g_{4}}\left\{M^{2}_{PL}\left[R_{(4)}-P\int^{+\pi R}_{-\pi R}dy \frac{4(3e^{2\beta y}+3\lambda^{2}e^{-2\beta y}-2\lambda)}{R^{2}(e^{\beta y}+\lambda
 e^{-\beta y})^{5}}\right]+\frac{e_{4}}{b_{0}}(\partial_{\alpha}
\phi^{\mu})^{\dag}(\partial^{\alpha}\phi_{\mu})\left(\frac{\partial^{2}
K(\phi,\phi^{\dagger})}{\partial\phi^{\dagger}_{\mu}\partial\phi^{\nu}}\right)\right.\\ \left.\displaystyle
+\frac{e_{4}C(T,T^{\dagger})}{4\pi^{2}R^{2}b_{0}}
e^{\frac{K(\phi,\phi^{\dagger})}{M^{2}}}\left[\left(\frac{\partial W}{\partial
\phi_{\alpha}}+\left(\frac{\partial K(\phi,\phi^{\dagger})}{\partial
\phi_{\alpha}}\right)\frac{W}{M^{2}}\right)^{\dag}\left(\frac{\partial^{2}K(\phi,\phi^{\dagger})}
{\partial\phi^{\alpha}\partial\phi^{\dagger}_{\beta}}\right)^{-1}
\left(\frac{\partial W}{\partial
\phi^{\beta}}+\left(\frac{\partial K(\phi,\phi^{\dagger})}{\partial
\phi^{\beta}}\right)\frac{W}{M^{2}}\right)-3\frac{|W|^{2}}{M^{2}}\right]\right\}
\\ \displaystyle=\frac{M^{2}_{PL}}{2}\int d^{4}x \sqrt{g_{4}}\left[R_{(4)}-P\int^{+\pi R}_{-\pi R}dy \frac{4(3e^{2\beta y}+3\lambda^{2}e^{-2\beta y}-2\lambda)}{R^{2}(e^{\beta y}+\lambda
 e^{-\beta y})^{5}}+\left(\frac{\partial^{2}
K(\phi,\phi^{\dagger})}{\partial\phi^{\dagger}_{\mu}\partial\phi^{\nu}}\right)(\partial_{\alpha}\phi^{\mu})^{\dag}(\partial^{\alpha}\phi_{\nu})-QV_{F}
\right].\end{array}\ee
where $P=\frac{2M^{3}_{5}\beta b^{6}_{0}}{M^{2}_{PL}R^{5}},~~Q=\frac{C(T,T^{\dag})}{4\pi^{2}R^{2}}$, $M_{PL}=M_{4}=\sqrt{\frac{e_{4}}{b_{0}}}=\sqrt{\frac{6e_{(5)}}{\lambda}}
=M^{3/2}_{5}\sqrt{V_{EXTRA}}=\frac{1}{\kappa_{4}}=\sqrt{\frac{6}{\lambda}}\frac{1}{\kappa^{2}_{5}}$, $\lambda=\frac{\Lambda_{5}b^{4}_{0}}{24R^{2}}$ and the compactification volume of the extra dimension $V_{EXTRA}=\frac{3M^{3}_{5}}{4\pi\lambda}$. Here
$C(T,T^{\dagger})$  represents an arbitrary
function of $T$ and $T^{\dagger}$. So eqn(\ref{ast8}) explicitly shows that the theory is reduced to  an effective $N=1,D=4$ SUGRA theory.
For a general physical situation of $N=1, D=4$
supergravity in the brane where the F-term potential on the brane
defined earlier is modified as
 \cite{nills},\cite{Antonio}
\be\label{vf}
V_{F}=\exp\left(\frac{K(\phi,\phi^{\dagger})}{M^{2}}\right)\left[\left(\frac{\partial W}{\partial
\Psi_{\alpha}}+\left(\frac{\partial K}{\partial
\Psi_{\alpha}}\right)\frac{W}{M^{2}}\right)^{\dag}\left(\frac{\partial^{2}K}{\partial\Psi^{\alpha}\partial\Psi^{\dagger}_{\beta}}\right)^{-1}\left(\frac{\partial W}{\partial
\Psi^{\beta}}+\left(\frac{\partial K}{\partial
\Psi^{\beta}}\right)\frac{W}{M^{2}}\right)-3\frac{|W|^{2}}{M^{2}}\right].\ee
  Here $\Psi^{\alpha}$ is the chiral superfield and $\phi^{\alpha}$
be the complex scalar field. From now on the inflaton field $\phi$ appears to be
4-dimensional as demonstrated earlier. Consequently for effective $N=1,D=4$ SUGRA eqn(\ref{em1}) and eqn(\ref{eqmos}) reduces to 
$T_{\alpha\beta}=\frac{g_{\alpha\beta}e_{4}V_{F}}{2\pi b_{0}R}$ and
$\partial_{\beta}\left(\sqrt{g_{4}}\partial^{\beta}\phi\right)+QV^{'}_{F}(\phi)=0.$
In this context we assume that
the K$\ddot{a}$hler potential is dominated by the leading order
term (first term) in canonical basis of the series representation 
i.e. $K= \sum_{\alpha}\phi^{\dagger}_{\alpha}\phi^{\alpha}$.
The superpotential in eqn(\ref{vf}) is given by 
$W=\sum^{\infty}_{n=0}D_{n}W_{n}(\phi^{\alpha})$
 with the constraint $D_{0}=1$. Here $W_{n}(\phi^{\alpha})$ is
a holomorphic function of $\phi^{\alpha}$ in the complex plane. 
Consequently in the canonical basis eqn(\ref{ast8}) takes the following form
\be\label{astre}S=\frac{M^{2}_{PL}}{2}\int d^{4}x \sqrt{g_{4}}\left[
R_{(4)}-P\int^{+\pi R}_{-\pi R}dy \frac{4(3e^{2\beta y}+3\lambda^{2}e^{-2\beta y}-2\lambda)}{R^{2}(e^{\beta y}+\lambda
 e^{-\beta y})^{5}}+(\partial_{\alpha}\phi^{\mu})^{\dag}(\partial^{\alpha}\phi_{\mu})-QV_{F}
\right],\ee
where
the F-term potential can be recast as ($V_{D} = 0 \Leftrightarrow
U(1)$ gauge interaction is absent) \cite{M.C.Bento}\be\label{totalpot}
V=V_{F}=\exp\left[\frac{1}{M^{2}}\sum_{\alpha}\phi^{\dagger}_{\alpha}\phi^{\alpha}\right]\left[\sum_{\beta}\left|\frac{\partial
W}{\partial \phi_{\beta}}\right|^{2}-3\frac{|W|^{2}}{M^{2}}\right].\ee
Now we expand the slowly varying inflaton potential derived from
F-term around the value of the inflaton field where the quantum
fluctuation is governed by, $\phi \rightarrow \tilde{\phi}+\phi$,
($\tilde{\phi}$ being the value of the inflaton field where
structure formation occurs) and by imposing $Z_{2}$ removing all odd order term responsible for
gravitational instabilities the required renormalizable inflaton potential turns
out to be \cite{Adams} $V=\Delta^{4}\sum^{2}_{m=0}C_{2m}\left(\frac{\phi}{M}\right)^{2m}$,
with another constraint $C_{0}=1$. The mass term decides the steepness of the potential.
 Absence of this term indicates that process is slow which is compensated by brane
 tension in the braneworld scenario \cite{riotto}. For the phenomenological purpose this specific choice 
is completely viable. But to incorporate thermal
history of the universe leading to  reheating and baryogenesis
among others we need to  perform the one loop corrected finite temperature
extension \cite{Narlikar} of our model. Now translating
the momentum integral within a specified cut-off ($\Lambda$) the effective potential turns out to be
\be\label{litl}
V(\phi)=\Delta^{4}+\frac{g}{4!}\phi^{4}+\frac{g^{2}\phi^{4}}{(16\pi)^{2}}
\left[\ln\left(\frac{\phi^{2}}{\Lambda^{2}}\right)-\frac{25}{6}\right]+O(\lambda^{3}),\ee
where the coupling constant $g=\frac{24\Delta^{4}C_{4}}{M^{4}}$ (Here $C_{4}$ is a tree level constant) which 
\cite{ryder} is, in general,  defined as 
$g(M)=\frac{d^{4}V(\phi)}{d\phi^{4}}|_{\phi=M}=g+\frac{g^{2}}{(8\pi)^{2}}\left[6\ln(\frac{M^{2}}{\Lambda^{2}})\right]+O(g^{3})$
so that the general expression
 for the effective potential  in terms of all finite physical parameters is given by
 \be\label{hhui}
V(\phi)=\Delta^{4}+\frac{g(M)}{4!}\phi^{4}+\frac{g^{2}(M)\phi^{4}}{(16\pi)^{2}}\left[\ln(\frac{\phi^{2}}{M^{2}})-\frac{25}{6}\right]+O(g(M)^{3}).\ee
which is the Coleman Weinberg potential \cite{Coleman}
,\cite{Landau}. 
After substituting the expression for $g$ in terms of $C_{4}$ the one loop corrected potential can be expressed as 
\be\label{post}V(\phi)=\Delta^{4}\left[1+\left(D_{4}+K_{4}\ln\left(\frac{\phi}{M}\right)\right)
\left(\frac{\phi}{M}\right)^{4}\right],\ee where 
$K_{4}=\frac{9\Delta^{4}C^{2}_{4}}{2\pi^{2}M^{4}} , ~~
D_{4}=C_{4}-\frac{25K_{4}}{12}$. This is precisely the potential eqn(\ref{opdsa}) mentioned in inflationary model building in the 
present paper.



\end{document}